# RESEARCH ARTICLE

EARTH SCIENCES

## Oldhamite: a new link in upper mantle for C–O–S–Ca cycles and an indicator for planetary habitability

Yuegao Liu [1,2,†], I-Ming Chou[1], Jiangzhi Chen[1,2,†], Nanping Wu[1], Wenyuan Li[3], Leon Bagas[3], Minghua Ren[4], Zairong Liu[1], Shenghua Mei[1,2,*] and Liping Wang[5,*]

[1]CAS Key Laboratory for Experimental Study under Deep-sea Extreme Conditions, Institute of Deep-sea Science and Engineering, Chinese Academy of Sciences, Sanya 572000, China; [2]Hainan Deep-Sea Technology Innovation Center, Sanya 572000, China; [3]Xi'an Center of Geological Survey, China Geological Survey, Xi'an 710054, China; [4]Department of Geoscience, University of Nevada, Las Vegas, Las Vegas, NV 89154, USA and [5]Academy for Advanced Interdisciplinary Studies, Southern University of Science and Technology, Shenzhen 518055, China

*Corresponding authors. E-mails: mei@idsse.ac.cn; wanglp3@sustech.edu.cn

†Equally contributed to this work.

Received 6 December 2022; Revised 17 May 2023; Accepted 24 May 2023

## ABSTRACT

In the solar system, oldhamite (CaS) is generally considered to be formed by the condensation of solar nebula gas. Enstatite chondrites, one of the most important repositories of oldhamite, are believed to be representative of the material that formed Earth. Thus, the formation mechanism and the evolution process of oldhamite are of great significance to the deep understanding of the solar nebula, meteorites, the origin of Earth, and the C–O–S–Ca cycles of Earth. Until now, oldhamite has not been reported to occur in mantle rock. However, here we show the formation of oldhamite through the reaction between sulfide-bearing orthopyroxene and molten CaCO₃ at 1.5 GPa/1510 K, 0.5 GPa/1320 K, and 0.3 GPa/1273 K. Importantly, this reaction occurs at oxygen fugacities within the range of upper-mantle conditions, six orders of magnitude higher than that of the solar nebula mechanism. Oldhamite is easily oxidized to CaSO₄ or hydrolysed to produce calcium hydroxide. Low oxygen fugacity of magma, extremely low oxygen content of the atmosphere, and the lack of a large amount of liquid water on the celestial body's surface are necessary for the widespread existence of oldhamite on the surface of a celestial body otherwise, anhydrite or gypsum will exist in large quantities. Oldhamites may exist in the upper mantle beneath mid-ocean ridges. Additionally, oldhamites may have been a contributing factor to the early Earth's atmospheric hypoxia environment, and the transient existence of oldhamites during the interaction between reducing sulfur-bearing magma and carbonate could have had an impact on the changes in atmospheric composition during the Permian–Triassic Boundary.

## INTRODUCTION

The alkaline- and alkaline-earth sulfides such as oldhamite (CaS) are extremely rare in terrestrial rocks. To date, only two studies have reported the potential occurrence of oldhamite in natural terrestrial rocks. One is in volcanic glass from the Arteni massif [1], and the other is from an impactite [2]. The is no report about the existence of oldhamite in the mantle and during the mantle–crust interaction process. However, oldhamite is potentially abundant in hollows and pits on Mercury's surface [3,4], which is closest to the Sun among the eight planets in the solar system, and it is a common mineral in enstatite chondrites [5] and aubrites (enstatite achondrites) [6] (Table 1), which are believed to be formed near the center of the solar nebula within the orbit of

Mercury [6,7]. In both cases, the highly reducing conditions with an oxygen fugacity well below IW-2.7 (IW = iron–wüstite redox buffer, in lg(o₂) stabilize oldhamite [8–10]. It seems that the initial formation of most oldhamites was in the region close to the Sun, at Mercury's surface or within the orbit of Mercury. At present, there is no discussion about the origin of oldhamite on Earth, and previous studies have focused on the origin of this mineral in enstatite chondrites and aubrite. There are currently two views regarding this issue.

Most scholars hold the view that oldhamite is a product of condensation of the solar nebula gas. Laboratory smoke experiments demonstrate that pure CaS condenses from vapor phases of calcium

**Table 1.** Proportional fractions of oldhamite in different enstatite meteorites.

| Classification | Group | Name | Proportion fraction of oldhamite | References |
|---|---|---|---|---|
| Enstatite chondrite meteorite | EL | Daniel's Kuil | 0.86 wt% | [11] |
| | | Khairpur | 0.34 wt% | [11] |
| | | Northwest Africa 1910 | 0.7 vol% | [12] |
| | | PCA91020 | n.d. | [13] |
| | | ALHA81021 | n.d. | [13] |
| | EH | ALHA77295 | 0.64 vol% | [14] |
| | | Sahara 97 072 | 1.12 vol% | [14] |
| | | LEW88180 | n.d. | [13] |
| | | EET 87 746 | ~3 wt% | [13] |
| | | Abee | 0.1−9 wt% | [15] |
| Aubrite (enstatite achondrite meteorite) | – | Bustee | 30 vol% | [6] |
| | | Norton County | 0.08−0.6 wt% | [16] |

EL, low Fe and siderophile group enstatite chondrite meteorite; EH, high Fe and siderophile group enstatite chondrite meteorite. Phases listed as n.d. were not detected in that sample.

and sulfur [17]. This supports a solar nebula gas origin and, according to first-principles calculations at equilibrium, oldhamite is more easily enriched in light Ca isotopes than other solid minerals. In contrast, condensed Ca-bearing minerals from nebula gas are enriched in heavy Ca isotopes relative to the residual gaseous Ca [18]. Oldhamite in enstatite chondrites is isotopically heavier than coexisting silicate materials, supporting the solar nebula gas origin. However, sulfur isotope data do not support the solar nebula gas origin. The correlation between $\Delta^{33}S$ and $\Delta^{36}S$ of some enstatite chondrites does not follow the trends of photochemistry in the solar nebula with $\Delta^{36}S = -2.98\Delta^{33}S$ [19] and of cosmic-ray spallation with $\Delta^{36}S = 8\Delta^{33}S$ [20].

Some scholars argue that the oldhamite in enstatite chondrites and aubrite is of igneous origin rather than the solar nebula gas origin [16,21]. Textural evidence includes apparent primary igneous grain boundaries between oldhamite and forsterite, and the presence of round, droplet-like Mn−Fe−Mg−Cr−Na sulfide inclusions within oldhamite, which appear to represent an immiscible sulfide liquid [16,21].

At the pressure (*P*)−temperature (*T*) conditions compatible with Earth's upper mantle and lower crust, the reaction experiments between sulfide-bearing orthopyroxenite and calcium carbonate were conducted in a multi-anvil cubic apparatus and a piston-cylinder press in this study. The experimental result shows that oldhamite can exist in the mantle and during the mantle–crust interaction process. Furthermore, through thermodynamic calculations and comparing the oxygen-fugacity values of the atmosphere and magma of different planets in the solar system, we infer that whether a large amount of CaS or CaSO₄ appears on the surface of a planet is closely related to the oxygen fugacity of the planetary magma and atmospheric composition. This paper could shed light on the formation of alkaline-earth metal sulfides, the origin of enstatite chondrites, C–O–S–Ca cycles, the mantle metasomatism mechanism, the crustal contamination process of mantle-derived magma and planetary habitability.

## RESULTS

### High *P*–*T* experiments

Here we show that oldhamite forms under conditions compatible with Earth's upper mantle and lower crust through the reaction between sulfide-bearing orthopyroxenite and CaCO₃ at 1.5 GPa/1510 K, 0.5 GPa/1320 K, and 0.3 GPa/1273 K in a graphite-lined Au₇₅Pd₂₅ capsule (Fig. 1). Oldhamite was observed in the central reaction zone of recovered samples (Fig. 1D and E and Supplementary Fig. S5). Hereafter we refer to this formation process as the sulfide–magma–calcite (SMC) interaction. In the absence of CaCO₃, the produced partial melt from orthopyroxenite under these three *P*–*T* conditions are basaltic melts, among which the melts produced under 1.5 GPa/1510 K and 0.5 GPa/1320 K are high-Mg basaltic melts (Fig. 1B) (with SiO₂ = 54.5−54.9 wt%, MgO = 9.54−10.19 wt%; Supplementary section 3 and Supplementary Table S3).

### Determination of oxygen-fugacity environment for the stable existence of oldhamite

In natural terrestrial samples, the occurrence of CaS is generally constrained by the following reactions:

$$CaS_{(s)} + 2O_{2(g)} = CaSO_{4(s)}$$

$$(\Delta H_{298\,K} = -959.5\ kJ/mol) \qquad (1)$$

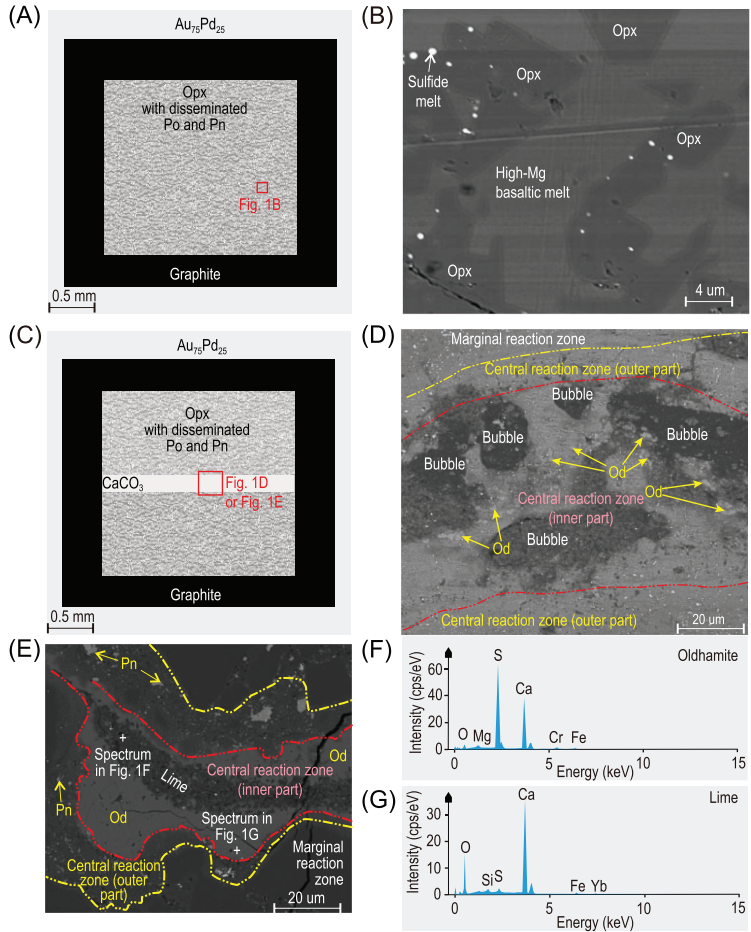

**Figure 1.** State and location of CaS generated by the interaction between sulfide-bearing magma and calcite (SMC). (A) Reaction chamber for the partial melting experiments of the Po–Pn-bearing orthopyroxenite; (B) the partial melting of the Po–Pn-bearing orthopyroxenite at 1.5 GPa/1510 K under scanning electron microscope (SEM); (C) reaction chamber for the contamination experiments between the Po–Pn-bearing orthopyroxenite and CaCO₃; (D) drop-shaped oldhamite in the inner part of the central reaction zone and disseminated Fe–Ni sulfide (bright white) at 0.5 GPa/1320 K; (E) oldhamite around lime at 0.3 GPa/1273 K; (F) the EDX/SEM spectrum of oldhamite; (G) the EDX/SEM spectrum of lime. Opx, orthopyroxene; Po, pyrrhotite; Pn, pentlandite; Od, oldhamite.

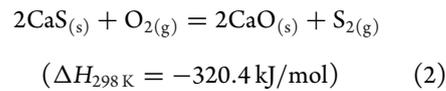

$$2CaS_{(s)} + O_{2(g)} = 2CaO_{(s)} + S_{2(g)}$$

$$(\Delta H_{298\,K} = -320.4\,\text{kJ/mol}) \qquad (2)$$

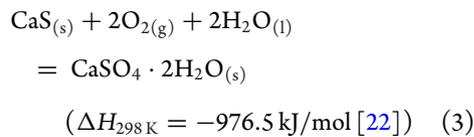

$$CaS_{(s)} + 2O_{2(g)} + 2H_2O_{(l)}$$
$$= CaSO_4 \cdot 2H_2O_{(s)}$$
$$(\Delta H_{298\,K} = -976.5\,\text{kJ/mol [22]}) \qquad (3)$$

As seen from these equations, CaS can be easily oxidized to form $CaSO_4$ or $CaSO_4 \cdot 2H_2O$. Thus, the stable existence of CaS depends mainly on oxygen fugacity. In order to quantitatively calculate the oxygen-fugacity boundary where CaS can exist stably, this paper defines two oxygen buffers for the first time: OA buffer and OLS buffer. The oxygen fugacity at the oldhamite–anhydrite equilibrium (Equation (1), named the OA buffer) and the oldhamite–lime–sulfur equilibrium (Equation (2), named the OLS buffer) can be determined by Equations (4) and (5), respectively:

$$\lg fo_2 = 2.19144 + 1.09305 \times 10^{-4}T - 25137/T$$
$$- 1551.42/T^2 + 1.5305 \times 10^7/T^3$$
$$+ 0.04777P/T + 2.7838 \lg T \qquad (4)$$

$$\lg fo_2 = -21.1162 + 3.65342 \times 10^7/T^3$$
$$- 6205.07/T^2 + (-16237.94$$
$$- 0.11450P)/T + 0.43722 \times 10^{-3}T$$
$$+ 11.13544 \lg T + \lg fs_2 \qquad (5)$$

where $P$ is the pressure in bar and $T$ is the temperature in K. The detailed process for the quantitative formula calculation of these two buffers is listed in Supplementary sections 5.2 and 5.3. The $T$–$\lg fo_2$ curves of these two buffers are shown in Fig. 2. At 0.5 GPa and 1320 K, OA = FMQ+2.21 = IW+6.05 ($\lg fo_2 = -7.83$) and OLS = FMQ−0.52 = IW+3.30 ($\lg fo_2 = -10.57$) (Fig. 2). If the oxygen-fugacity value is lower than OLS, oldhamite is stable. On the contrary, the oxygen-fugacity value of the anhydrite stable field is higher than OA.

## DISCUSSION

### New formation mechanism of oldhamite

The partial melts from orthopyroxenite in our experiments are basaltic melts or high-Mg basaltic melts. This is consistent with the partial melting process of mantle pyroxenites that in part produced mid-ocean ridge basalts (MORBs) and Hawaiian shield basalts [35,36]. It is worth noting that similar high-Mg basaltic melt is the parent magma of some magmatic Cu–Ni–Pt deposits in orogenic belts or related to mantle plumes [37]. These melts are derived from the mantle but usually interacted with crustal carbonate [38]. The oxygen fugacity in the graphite-lined noble metal capsule used in this study is about FMQ−2.1 [39] (FMQ= fayalite–magnetite–quartz redox buffer, in $\lg fo_2$). Hence, our experiments simulate the reaction between basaltic and carbonate magma in the mantle, the metasomatism of the mantle by carbonate melts, and the reaction process between basaltic magma and crustal carbonate. Based on our findings, oldhamite can exist in the mantle under these conditions. In our run products, limes (CaO), which were formed by the decomposition

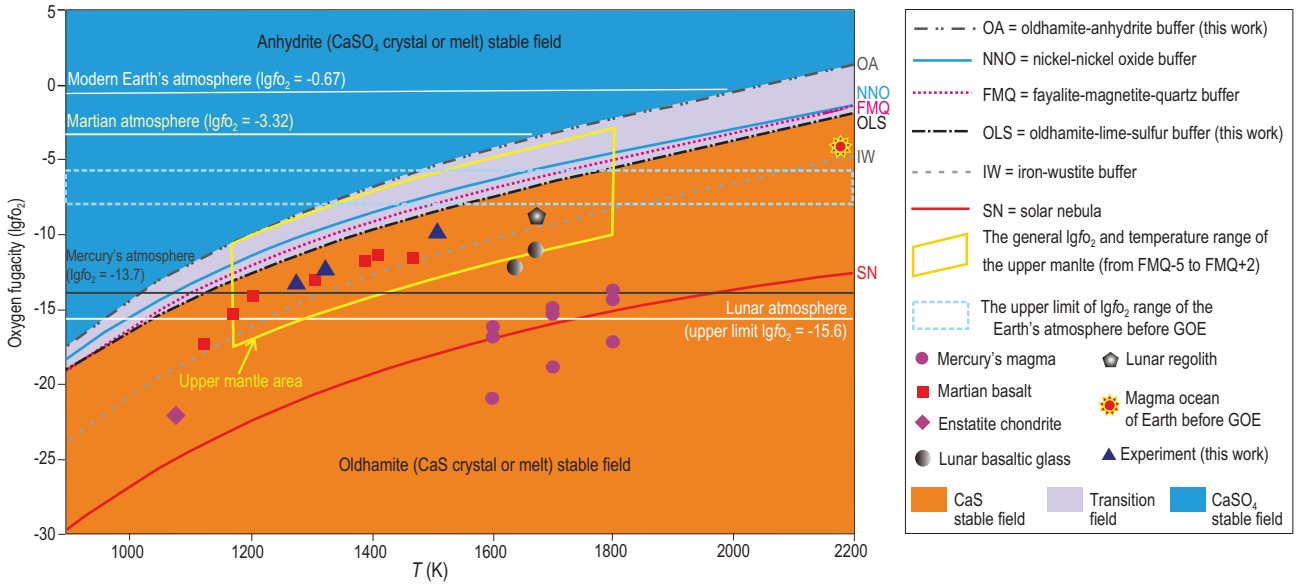

**Figure 2.** Representative oxygen-fugacity buffer at 0.5 GPa total pressure vs. temperature curves and the distributions of some natural and experimental samples. References: NNO is from Ref. [23], and FMQ and IW are from Ref. [24]; lg$f_{S_2}$ value of $-12$ (mean sulfur fugacity of black smoker [25]) was used during the OLS calculation. Values of solar nebula lg$f_{O_2}$ are calculated from the equation lg$f_{O_2} = -0.85-25\,664/T$, where $T$ is in K [26]. The lg$f_{O_2}$ values of enstatite chondrite, Martian basalt, the silicate melts on Mercury's surface, lunar basaltic glass, lunar regolith, and the magma ocean of early Earth are from Refs [8], [27], [28], [29], [30] and [31], respectively. The general lg$f_{O_2}$ and temperature range of the upper mantle are from Refs [32] and [33], respectively. The upper limit of lg$f_{O_2}$ range of Earth's atmosphere before GOE is from Ref. [34].

of $CaCO_3$, were observed in the central reaction zone (Fig. 1E). Pentlandite or pyrrhotite is present around oldhamite, and cavities formed by bubbles are present close to oldhamite (Fig. 1D and Supplementary Fig. S5). Hence, the most probable route for CaS formation is:

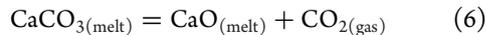

$$CaCO_{3(melt)} = CaO_{(melt)} + CO_{2(gas)} \quad (6)$$

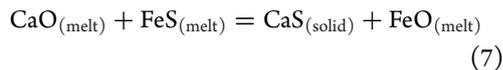

$$CaO_{(melt)} + FeS_{(melt)} = CaS_{(solid)} + FeO_{(melt)} \quad (7)$$

Here, we name this process the sulfide–magma–calcite interaction model (SMC model).

The SMC model is different from the previous genetic model in three ways. The first difference is the formation location. Previous researchers believed that oldhamite exists in the enstatite meteorite, on the surface of Mercury, and in the lunar regolith (Fig. 3) [3,5,40], but we proved that it could be stable in the upper mantle. It is reasonable that it could be formed in the interior of other terrestrial planets with magmatic activities. This expands the distribution range of oldhamite in the solar system (Fig. 3). The second difference is that the formation process by the SMC interaction in this study is totally different from the nebula gas mechanism. Besides, oldhamite in lunar regolith is considered to be formed by the amalgamation of vapor phases of Ca and S produced during meteorite impact [40]. This is also completely different from the SMC model. The third difference is about oxygen fugacity. Former researchers thought that an oxygen fugacity below IW–2.7 is necessary for the stable existence of oldhamite in the solar nebula model [8–10]. However, our experiment and calculation results show that the oldhamite can be stable below IW+3.3 (Fig. 2), which is six orders of magnitude higher than the limit in the solar nebula mechanism.

## The preservation conditions of oldhamite on the surface of a celestial body

Our results support the presence of oldhamite in the mantle, but why is it hard to be found on Earth's surface? As the thermal expansion coefficient of oldhamite is high with a value of $4.03 \times 10^{-5}\,K^{-1}$, the temperature is an important factor affecting whether oldhamite can maintain its crystal form [41]. But the oldhamite crystal is stable at Mercury's surface with a high surface temperature of $\leq 723.15$ K [41], which is much higher than that of Earth's surface. Thus, the surface-temperature factor is not the reason why it is hard to find oldhamite on Earth's surface. A closer look at the redox conditions of the hosting magma and atmosphere is required. Oldhamite (including crystal and melt) is stable when lg$f_{O_2}$ is below the OLS buffer, but cannot exist when lg$f_{O_2}$ is above the OA buffer (Fig. 2). The lg$f_{O_2}$ values of Mercury's sur-face magma range from IW−5.4 to −10.5 ($\lg fo_2 = −13.7$ to $−20.8$) at 1600−1800 K (Fig. 2), and the $\lg fo_2$ value of Mercury's atmosphere is −13.7 (Supplementary section 5.1). These values are much below the OLS buffer (Fig. 2), permitting the conservation of oldhamite. That is, the widespread presence of oldhamite on the surface of a celestial body indicates an extremely anoxic environment. Oldhamite has also been found in the lunar regolith, which is a weathering product of rocks formed by meteorite impacts [40]. The oxygen fugacity of the lunar basaltic glass and regolith are ∼IW−1.7 ($\lg fo_2 = −11.7$) and IW+1 ($\lg fo_2 = −9.1$) at 1673 K, respectively [29,30], and the upper-limit $\lg fo_2$ value of the lunar atmosphere is −15.6 (Supplementary section 5.1). These oxygen fugacities are lower than the OLS buffer, which meets the conditions for the presence of oldhamite. On the other hand, as the $\lg fo_2$ value of −4.32 of the Martian atmosphere is higher than OLS, no oldhamite but some sulfates have been found on Mars, despite the $\lg fo_2$ of Martian magma being lower than OLS (Fig. 2). It means that the extremely low oxygen content in the atmosphere is necessary for the preservation of oldhamite, which is unsuitable for most known organisms to survive. Interestingly, Earth takes an intermediate position between Mercury and Mars (though closer to Mars). The average $\lg fo_2$ values for arc basalts (FMQ+0.96), ocean island magma (FMQ+0.82), and most basalt related to mantle plumes on Earth (FMQ+0.1) are higher than OLS [42]. Moreover, the $\lg fo_2$ value of −0.67 of Earth's atmosphere is much higher than OA (Fig. 2). The habitable surface conditions for humans on Earth do not support the preservation of oldhamite.

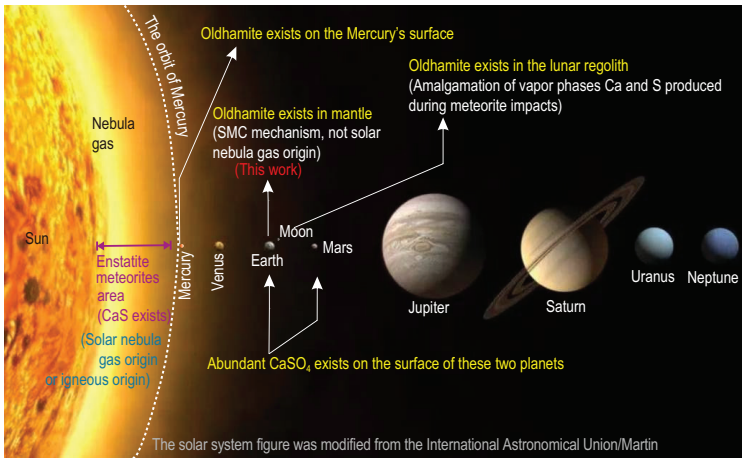

**Figure 3.** The known main distribution of oldhamite (CaS) in the solar system.

Moreover, in aqueous solution, CaS can be hydrolysed to form calcium hydroxide. The abundant presence of liquid water on Earth's surface is also detrimental to the preservation of CaS. On the contrary, although there is evidence to suggest the presence of water ice in permanent shadow areas within polar craters of Mercury and the Moon [43,44], liquid water has not been reported to occur on the surfaces of these two celestial bodies. Thus, the lack of a large amount of liquid water on the surface perhaps is necessary in the preservation of oldhamites on a celestial body's surface.

## The possible link between oldhamite and anhydrite in black smokers

Most of the oxygen fugacities of the upper mantle are located in the oldhamite stable field (Fig. 2), but it is hard to be found on Earth's surface. A reasonable explanation is the existence of an interface where CaS is converted to $CaSO_4$ between the upper mantle and the surface. On Earth, the MORBs are characterized by a redox state of FMQ−0.41 ± 0.43 (Supplementary Fig. S6) [45], which is close to FMQ−0.52, the upper-limit oxygen fugacity for the stable existence of oldhamite. There is a large amount of anhydrite and gypsum in the mid-ocean ridges black smoker system. Thus, the interface where CaS is converted into $CaSO_4$ may exist near or at the solidified MORBs (Supplementary Fig. S6). In the mantle underneath mid-ocean ridges, the conditions for the formation of CaS including carbonate magma and sulfide-bearing magma sometimes can be met. The super-deep diamond research revealed that the oxygen fugacity of the bottom of the mantle transition zone can be IW−6.7 [46], which is much lower than OLS. Generally, at a depth of ∼160−170 km, the diamond is expected to convert into graphite at an oxygen fugacity of >FMQ−2 (Supplementary Fig. S6) [47,48]. Redox melting [C (graphite) + $2Fe_2O_3$ (melt) + $O^{2−}$ (melt) = 4FeO + $CO_3^{2−}$ (in the melt)] takes place at a depth of ∼120−150 km with an oxygen fugacity of ∼FMQ−1.6 (Supplementary Fig. S6) [48]. The carbonate melt produced by the redox melting will then ascend as a flux into the overlying mantle [49]. The interaction between carbonate melt and sulfide-bearing basaltic magma could happen, providing there are conditions for the formation of oldhamite.

The $\delta^{34}S_{V-CDT}$ value of anhydrite in the mid-ocean ridges black smoker system gradually decreases from the seawater's value of 20 ± 1 to 3.4‰, as observed in the 1.8-km-deep drill hole in a middle-ocean ridge [50]. The $\delta^{34}S_{V-CDT}$ value of 3.4‰, which is much lower than that of modern seawater, is considered to have resulted from the oxidation of low-$\delta^{34}S$ sulfide to sulfate in the MORB [51]. Oldhamite is easily oxidized due to the large negative $\Delta H_{298\,K}$ value for Equation (1). Thus, the oxidation of CaS to $CaSO_4$ could be a viable route for the formation of sulfate. On the other hand, oldhamite is more easily enriched in light Ca isotopes than other Ca-bearing minerals [18,52]. The dissolution of $CaSO_4$ that has experienced the CaS–$CaSO_4$ isotopic fractionation into the hydrothermal fluids at the mid-ocean ridges is expected to increase the $\delta^{44/40}$Ca value of the fluids relative to the host-rocks, which, indeed, has been observed [53]. Thus, the oxidation of CaS to $CaSO_4$ offers a viable alternative for the origin of anhydrite in the mid-ocean ridges black smoker system, though a lot of gypsum in the mid-ocean ridge is due to the decrease in the solubility of Ca and $SO_4^{2-}$ in seawater with increasing temperature [54]. The new sulfate-formation mechanism, the CaS oxidation model in this paper, can explain some special Ca–S isotope characteristics, which is a supplement to the formation process of sulfate on planets.

## The influence of oldhamite on atmospheric composition before the Great Oxidation Event

Earth's deep interior holds the key to habitability [55–57]. Oldhamites in Earth may affect the atmospheric composition. Many pieces of evidence show that Earth's atmosphere before the Great Oxidation Event (GOE) is initially free of $O_2$ [58,59]. Around 2.46−1.85 billion years ago (Ga), oxygen levels rose from $<10^{-7.1}$–$10^{-5.1}$ that of present atmospheric levels (PAL) ($lg fo_2 < -7.6$ to $-6.0$, Fig. 2) to $10^{-4.6}$–$10^{-2.0}$ PAL ($lg fo_2 = -5.3$ to $-2.7$) [34], known as the GOE. To explore the reasons for the anoxic or oxygen-free feature of early Earth's atmosphere, we need to clarify the characteristics of the initial materials that were used to build Earth. Enstatite chondrites, one of the most important repositories of oldhamite (Table 1), are the most-reduced meteorites and have similar isotopic composition to terrestrial rocks, so are often considered to be representative of the material that formed Earth [60,61]. That is, early Earth probably contains a significant amount of oldhamite. The Earth–Moon precursor materials with oldhamite enrichment could well explain the Sm/Nd ratio and Nd isotope features of Earth and the Moon [62]. The upper limit of the $lg fo_2$ range of Earth's atmosphere before GOE is from −7.6 to −6.0 (Fig. 2) [34], which partially overlaps with the oldhamite stable field. Besides, the $lg fo_2$ range of magma ocean of early Earth is IW+0.5 at 2173 K [31] (Fig. 2). Thus, both the atmospheric oxygen fugacity and the magmatic oxygen fugacity met the conditions for the existence of oldhamite, which would inhibit the generation of free oxygen due to the oxidation of oldhamite to $CaSO_4$ (Equation (1)).

The mass of Earth is $\sim5.97 \times 10^{24}$ kg [63]. The mass fraction of oldhamite in early Earth is estimated to be 0.072%−0.127% (Supplementary section 8). Thus, we can infer the mass of oldhamite in early Earth is $\sim4.30 \times 10^{21}$−$7.58 \times 10^{21}$ kg. If all these oldhamites are converted into $CaSO_4$, they will consume $3.82 \times 10^{21}$−$6.74 \times 10^{21}$ kg $O_2$. The total $O_2$ amount of modern Earth's atmosphere is only $\sim1.246 \times 10^{18}$ kg, 23% of the modern atmosphere's mass, where the atmosphere's mass is equal to $5.148 \times 10^{18}$ kg [64]. It is clear that the potential influence of oldhamite on atmospheric oxygen content is enormous. After GOE, the $lg fo_2$ of Earth's atmosphere at $\sim1.85$ Ga ranges from −5.3 to −2.7 [34], which is higher than OLS. In this oxygen-fugacity condition, some $CaSO_4$ are expected to appear on Earth after GOE. Calcium sulfate ($CaSO_4$) layer has indeed been observed in the 2.2-Ga sedimentary rocks in the Yerrida rift basin of Western Australia [65]. Thus, oldhamite could have played a role in Earth's anoxic or oxygen-free atmosphere before GOE.

## The influence of oldhamite on atmospheric composition during the Permian–Triassic boundary

The Permian–Triassic boundary (PTB) mass extinction was the most severe biotic crisis in the past 500 million years [66,67]. During the PTB mass extinction period, a sharp increase in atmospheric $CO_2$ content and a decrease in atmospheric $O_2$ content occurred. The atmospheric $CO_2$ concentration at the PTB is estimated to have been $3314 \pm 1097$ ppm, which is more than double the Permian average and is $12 \pm 4$ times that of the current atmosphere [68]. The atmospheric oxygen underwent a very sharp drop from 30% to 15% (volume fraction) at the PTB [69]. The formation of 1 mole of CaS is accompanied by the production of 1 mole of $CO_2$ (Equations (6) and (7)) and the oxidation of 1 mole of CaS to sulfate will consume 2 moles of $O_2$ (Equation (1)). The reaction between mantle magma and crustal rocks in the Siberian large igneous province (SLIP) is believed to be an important trigger for the above atmospheric composition change [70]. The SLIP is characterized by tholeiitic basalts [71], whose composition is similar to the composition of the initial melt of the partial melting of orthopyroxenite in this study (Supplementary Table S3). The magma of SLIP is characterized by a low oxygen fugacity with a value of FMQ–1.5 [71]. Moreover, some famous magmatic Cu–Ni sulfide ore deposits include the world's largest one formed at the SLIP, indicating the presence of some sulfide-rich magma. The above features meet the conditions for the formation of CaS. Thus, the intermediate effect of CaS, accompanied by $CO_2$ generation and $O_2$ consumption, can hardly be excluded during the interaction process between the upwelling of large-scale reducing S-bearing magma from mantle and the crustal carbonates in the SLIP.

## CONCLUSIONS

Two geological processes are likely to involve oldhamite as a transient phase, including mantle metasomatism by carbonate melts beneath the MORB region and crustal calcite contamination of mantle-derived magma during the formation of some magmatic Cu–Ni–PGE sulfide deposits at the Siberian large igneous provinces. Oldhamite is a plausible precursor for igneous Ca-sulfate in MORB. The formation and oxidation of oldhamite are accompanied by the production of $CO_2$ and the consumption of $O_2$, which have more or less influence on the atmospheric composition. Widespread existence of oldhamite on a planet's surface indicates a low oxygen fugacity for the magma of the planet, extremely low oxygen content in its atmosphere, and the lack of a large amount of liquid water on the planet's surface, which is not habitable.

## METHODS
### High *P–T* experiments

Experimental petrological methods are used to simulate the formation process of oldhamite in the mantle. The initial material is pyrrhotite-pentlandite-bearing orthopyroxenite and $CaCO_3$ powder. The mineral composition and chemical composition of starting materials are described in Supplementary section 1. Experiments were conducted at 0.5 GPa/1320 K and 1.5 GPa/1510 K, using a 2000-ton multi-anvil cubic apparatus at the University of Nevada, Las Vegas. Besides, the experiment under 0.3 GPa/1273 K was performed by using a piston-cylinder press at the Institute of Deep-sea Science and Engineering, Chinese Academy of Sciences. Sample powder was packed in a graphite crucible placed in an $Au_{75}Pd_{25}$ outer capsule with an outer diameter of 3 mm (Supplementary Figs S3A and S4). The mineral and melt composition analysis after the experiment was completed by using an electron microprobe. Please refer to Supplementary section 4 for the parameter design of the electron probe.

## The determination of oxygen fugacity

The oxygen fugacities at the oldhamite–lime–sulfur equilibrium (OLS buffer) and at oldhamite–anhydrite equilibrium (OA buffer) were obtained by using thermodynamic calculation. The detailed calculation processes are listed in Supplementary sections 5.2 and 5.3.

The concept of the oxygen fugacity of rocks has been deeply rooted in people's minds, but some researchers are not familiar with the oxygen fugacity of the atmosphere. Here, we explain the calculation process of the oxygen fugacity of modern Earth's atmosphere. Oxygen accounts for 21% of modern Earth's atmospheric volume, so the partial pressure of oxygen is 0.21 bars. Then, log (0.21) = −0.67. That is, the lg$f_{O_2}$ of modern Earth's atmosphere is −0.67 (Fig. 2). Similarly, the oxygen fugacities of Mercury's atmosphere, the Martian atmosphere, and the lunar atmosphere are also calculated in Supplementary section 5.1.

## The determination of isotope fractionation

Past researchers have calculated the polynomial fitting of the ratio of the reduced partition function for $^{44}Ca/^{40}Ca$ of oldhamite and anhydrite by using the static first-principles calculation [18,52]. Based on these results, the equilibrium Ca isotope fractionation between $CaSO_4$ and CaS is inferred. The inferred equation is listed in Supplementary section 6.

## DATA AVAILABILITY

All data are available in the main text or the Supplementary materials.

## SUPPLEMENTARY DATA

Supplementary data are available at *NSR* online.

## ACKNOWLEDGEMENTS

We thank Professor Oliver Tschauner from the University of Nevada, Las Vegas for many in-depth discussions. We are very grateful to Professor D.L. Kholstedt from the University of Minnesota-Twin City and Professor Y.G. Xu and Dr Z.Y. Luo from the Guangzhou Institute of Geochemistry, Chinese Academy of Sciences for their assistance in experimental petrology and sample preparation. Professor F. Huang and C. Zhou from the University of Science and Technology of China are acknowledged for their great help in isotope interpretation. Dr Nanfei Cheng from the Institute of Deep-sea Science and Engineering, Chinese Academy of Sciences helped a lot during the experiment in the piston-cylinder press. We acknowledge two anonymous reviewers and the Editorial Board for constructive suggestions.

## FUNDING

This work was supported by the Hainan Provincial Joint Project of Sanya Yazhou Bay Science and Technology City (2021CXLH0027 to Y.G.L.), the Central Guidance on Local Science and Technology Development Fund (ZY2021HN15 to S.H.M.), the Chinese Academy of Sciences (QYZDY-SSW-DQC029 and XDA22040501 to S.H.M.), the National Natural Science Foundation of China (41973055 and 42130109 to I.M.C), and the Major Science and Technology Infrastructure Project of Material Genome Big-Science Facilities Platform supported by Municipal Development and Reform Commission of Shenzhen (L.P.W.).

## AUTHOR CONTRIBUTIONS

Conceptualization: Y.G.L., I.M.C., S.H.M. Methodology: Y.G.L., J.Z.C., L.P.W., N.P.W., W.Y.L., Z.R.L., and M.H.R. Investigation: Y.G.L., J.Z.C., N.P.W., and L.P.W. Visualization: Y.G.L., I.M.C., J.Z.C., W.Y.L., L.B., and M.H.R. Supervision: L.P.W., S.H.M., I.M.C. Writing—original draft: Y.G.L., J.Z.C., L.P.W., L.B., S.H.M., and I.M.C. Manuscript editing: S.H.M., I.M.C., L.P.W., L.B., Z.R.L., and M.H.R.

*Conflict of interest statement.* None declared.